\documentclass[journal,comsoc]{IEEEtran}
\ifCLASSINFOpdf
 \usepackage[pdftex]{graphicx}

\else

\fi

\usepackage{cite}
\usepackage{amsmath,amssymb,amsfonts}
\usepackage{algorithmic}
\usepackage{graphicx}
\usepackage{textcomp}
\usepackage{subfigure}
\usepackage{threeparttable}
\usepackage{float}
\usepackage{epsfig}
\usepackage{epstopdf}
\usepackage{amsmath}
\usepackage{multirow}
\usepackage{color, soul}
\soulregister\cite7 
\soulregister\citep7 
\soulregister\citet7 
\soulregister\ref7 
\soulregister\pageref7 

\def\BibTeX{{\rm B\kern-.05em{\sc i\kern-.025em b}\kern-.08em
    T\kern-.1667em\lower.7ex\hbox{E}\kern-.125emX}}
\usepackage{makecell}
\usepackage[bookmarksnumbered=true, bookmarksopen=true]{hyperref}

\begin{document}

\title{Channel Modeling for UAV Communications: \\State of the Art, Case Studies, and Future Directions}
  \author{Zhuangzhuang~Cui,~Ke~Guan,~C\'esar~Briso-Rodr\'iguez,~Bo~Ai,~Zhangdui~Zhong,~and~Claude Oestges
  \thanks{Z. Cui, K. Guan, B. Ai and Z. Zhong are with Beijing Jiaotong University and Beijing Engineering Research Center of High-Speed Railway Broadband Mobile Communications.~C. Briso-Rodr\'iguez is with Universidad Polit\'ecnica de Madrid. C. Oestges is with Universit\'e Catholique de Louvain.}}


\maketitle

\begin{abstract}
As essential aerial platforms, unmanned aerial vehicles (UAVs) play an increasingly important role in broad wireless connectivity and high-data-rate transmission for future communication systems. Notably, various communication scenarios are involved in UAV communications, such as intercommunications between UAVs and communications with the ground user equipment, the cellular base station, and the ground station, to name a few. However, existing works mostly focus on a single communication scenario, a designated channel type, and a specific operating frequency, thus urgently requiring a comprehensive understanding of multi-scenario, multi-frequency, and multi-type UAV channels. This article pours attention into the essentials of corresponding air-to-air (A2A) and air-to-ground (A2G) channels in UAV communications. We first identify the latest key challenges of channel modeling for UAV communications. We then provide the state of the art for A2A and A2G channel properties and models based on extensive measurement campaigns. In particular, we conduct realistic case studies to further demonstrate critical channel characterizations and machine learning-based modeling methods. Last but not least, potential directions are widely discussed for paving the way towards more accurate and effective channel models for UAV communications.
\end{abstract}


\section{Introduction}
Incorporating unmanned aerial vehicles (UAVs) into wireless communications has attracted increasing interest during decades. From Release 15, the 3rd generation partnership project (3GPP) started the study on enhanced long-term evolution (LTE) support for aerial vehicles \cite{a11}. To further meet the emerging needs for UAVs in the fifth-generation (5G) communication system, the 3GPP Release 17 and 18 are working on the UAV connectivity, identification, and tracking under the topic of 5G enhancement for UAVs. Besides, UAV serving as an aerial base station (ABS) can assist or substitute a terrestrial base station (TBS) in specific situations. Thus, UAVs are playing an important role in a slice of arising communication systems. As shown in Fig.~1, various ground terminals can connect with UAVs, such as ground station (GS), user equipment (UE), vehicles, Internet-of-Thing (IoT) node, and TBS. In this article, we focus on typical UAV communication scenarios that mainly consist of three categories: UAV-to-UAV, UAV-to-BS, and UAV-to-GS/UE \cite{x2}. The key features of these communication scenarios are summarized as follows.
\begin{itemize}
\item \textbf{\emph{UAV-to-GS/UE:}} This refers to the command and control (C2) link between UAV and GS, operating in unlicensed bands such as 2.4 and 5~GHz. Subsequently, the UAV is considered an auxiliary or alternative of TBS, serving ground UEs as an ABS. Therefore, the operating bands on this link range from hundreds of Megahertz (MHz) to millimeter-wave (mmWave) bands, in which L-band for control and non-payload communication (CNPC) links and C-band for payload communication links are recommended by the International Telecommunication Union.
\item  \textbf{\emph{UAV-to-BS:}} To guarantee reliable beyond-line-of-sight (BLOS) communication links in long-distance flights, UAVs are expected to be aerial users of almost ubiquitous TBSs in cellular networks, which is known as cellular-connected UAV (CC-UAV) communications. Since the TBS is equipped with down-tilt directional sector antennas and deployed in fixed heights, which differs from the GS/UE, we category them as distinct scenarios.
\item  \textbf{\emph{UAV-to-UAV:}} For many use cases such as aerial relay and flying ad hoc network, intercommunications between UAVs are necessary. Compared with conventional vehicular communication, UAV-to-UAV communications exhibit more challenges since drones can fly with highly variable heights in the three-dimensional (3D) space, comparably, vehicles generally travel along with linear trajectory on the two-dimensional (2D) plane.
\end{itemize}

Regarding propagation channels, the air-to-air (A2A) and air-to-ground (A2G) channels are involved in the communication links mentioned above. Note that A2G channels include UAV-to-GS/UE and UAV-to-BS scenarios. It is undeniable that an accurate channel model is paramount for implementing reliable communication links. Thus, the article aims to investigate state-of-the-art channel models and to illustrate important factors that can affect UAV-related propagation channels.

It is known that popular channel models are usually statistical, deterministic, and geometry-based stochastic. Each modeling approach has its pros and cons. For example, the effectiveness of the ray-tracing-based deterministic modeling is highly dependent on the accuracy of the 3D reconstructed environment and corresponding electromagnetic properties of scatterers in the environment. For better generality, geometry-based stochastic channel models (GBSCMs) are under extensive study, however, the predefined geometry may deviate from the reality and thus reduce the accuracy. Due to specific measured frequencies and environments, measurements may lack favorable expansibility. However, numerous measurements have been conducted for plenty of frequencies, environments, and UAV altitudes, which are supportive of providing a comprehensive investigation. Thereby we study UAV-related propagation channels mainly based on channel measurements conducted in the literature and our previous work. More specifically, in the article, we are committed to answering the following questions:
\begin{itemize}
\item  What are the key challenges of channel modeling for UAV communications?
\item  What is the state of the art of UAV channel modeling, and what are the directions for future work?
\item  What are the impacts of frequency and environment on the channel characterization?
\item  Will machine learning (ML) be effective for modeling UAV channels?
\end{itemize}

\begin{figure}[!t]
  \centering
  \includegraphics[width=3.5in]{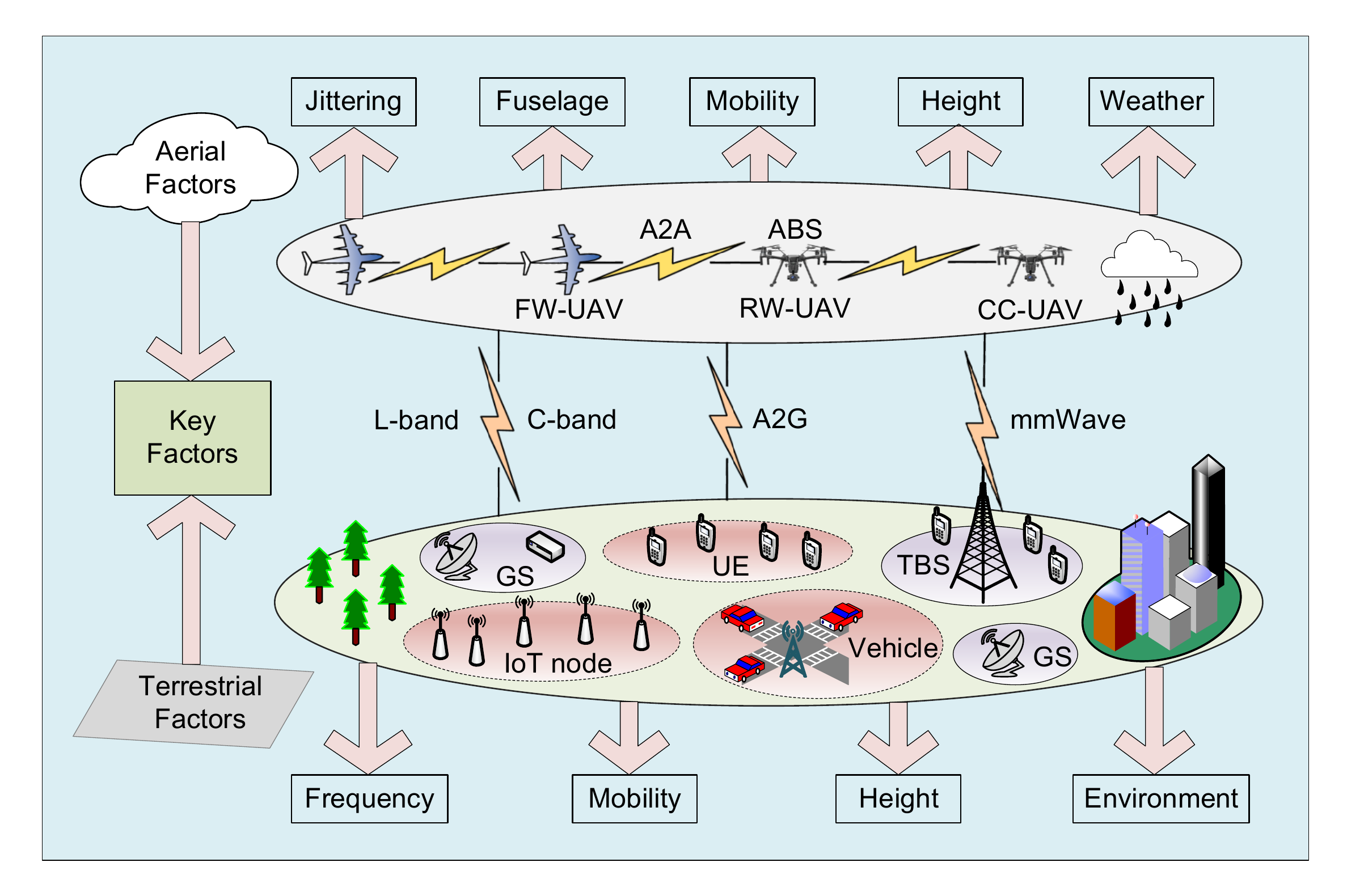}
  \caption{Schematic diagram of typical UAV-related communication scenarios.}
  \label{AG-CAN}
 \end{figure}
\section{Key Challenges of UAV Channel Modeling}
The special features of UAVs and their flights directly induce some key challenges in channel modeling, compared with traditional cellular or vehicular communications. As shown in Fig. 1, the key factors are not only related to external factors including frequency, environment, and weather, but also dependent on internal characteristics of UAVs and ground terminals. We highlight key challenges in detail as follows.

\textit{Variable heights:} Based on the flying altitude, aerial platforms can be divided into the high-altitude platform (HAP) with altitudes ranging between 17-32 km, and the low-altitude platform (LAP) with altitudes below 5~km. Notably, the Federal Aviation Administration (FAA) has stipulated that the maximum allowable altitude of drones weighing less than 55 pounds (25~kg) is 400~feet (121.92 m) above the ground level (AGL) \cite{f1}. Moreover, as the potential next regulatory limit, the 3GPP selected the maximum altitude as 300~m for LTE/5G-supported UAV communications. Thereby, a robust UAV communication system should adapt to different heights for both UAV and ground terminals, which requires researchers to accurately quantify the impact of altitude on channel characteristics. Specifically, the height-dependent channel models are indispensable \cite{x2}.

\textit{Changeable mobility:} UAVs can be roughly divided into the fixed-wing (FW) and rotor-wing (RW) kinds. The main difference lies in that RW-UAVs can hover in the air with zero speed, whereas FW-UAVs need to maintain the aerodynamic lift by keeping certain mobility. Overall, the mobilities of UAV and ground terminals such as vehicles are highly changeable. However, the high-speed mobility is adverse to reliable transmission, since propagation channels are vulnerable to suffer from spatio-temporal non-stationarity.


\textit{Flying fluctuations:} Extremely stable flights are impractical for any kind of UAVs. Thus, the impacts of fluctuating/jittering/wobbling flight of UAV on wireless channels should be carefully considered. The study in \cite{f2} has shown that a small variation in the UAV-UE distance led by unstable flight can cause a large phase offset of multipath components (MPCs), and thus severely affects the channel coherence time. Besides, it is meaningful to consider different flying states for UAVs, such as taxiing, en-route, takes-off, and landing, since these states may cause different levels of fluctuations.

\textit{Fuselage shadowing:} The shadowing effects due to the blockage in the flight may cause signal interruption. Obviously, the shadowing is highly related to the fuselage size and flying altitude of the UAV, as well as the antenna placement. Thus, in the large-scale fading model, these influences should be comprehensively considered.

\textit{Rain attenuation:} The rainfall, as a common weather condition, will lead to significant attenuation due to absorption and scattering for frequencies above 10~GHz. As an example, the rain attenuation was measured at 37.3-39.2 GHz with distances between 48-497 m in \cite{f3}. The results show that the attenuation ranges from 1.5 dB to 9 dB for different rainfall rates. It brings another challenge for UAV mmWave links.

\textit{Various environments and frequencies:} UAVs can fly in various environments, such as built-up, over-water, forest, mountain, and other remote or harsh environments. Since the density, height, distribution, and electromagnetic parameters of scatterers are distinct in these environments, multi-environment measurements for UAV channels are essential. Besides, the diversity of the frequency is also of significance for channel modeling. The frequency bands used in UAV communications vary from sub-6~GHz to mmWave bands, which requires more effort to characterize the impact of diverse bands on channel characteristics.

\section{State of The Art}
In this section, we first enumerate the recent progress of A2A and A2G channel modeling in terms of large-scale and small-scale statistics. Then, we thoroughly review channel models that were proposed in the literature. The state of the art not only provides a better understanding of UAV propagation channels but also inspires some potential directions to improve the deficiencies of existing works.
\begin{table*}[!tbp]
\centering
  \caption{A Summary of Representative A2G and A2A Channel Measurements and Statistics.} \label{table2}
  \begin{tabular}{|l|l|l|l|l|l|l|}
\hline
\textbf{Link} & \textbf{ Ref.}  &\textbf{UAV Altitude} &\textbf{Frequency Selectivity} & \textbf{Terminal}& \textbf{Environment} & \textbf{Channel Statistics} \\
    \hline
\multirow{5}{*}{A2G} & \cite{a1} & 580, 800~m &WB, 968, 5060~MHz & GS &  Over-water & $n$=1.5-2.2, $\sigma_s$=2.6-4.2~dB, PDP, SD, KF, RMS-DS \\
 \cline{2-7}
 & \cite{a2} &  504-924~m & WB, 968, 5060~MHz  & GS & Suburban, Near-urban & $n$=1.5-2.0, $\sigma_s$=2.6-3.2~dB, PDP, KF, RMS-DS, SIC \\
 \cline{2-7}
 & \cite{a4} &  4-16~m & WB, 3.1-5.3~GHz & GS  & Open-field, Suburban & $n$=2.5-3.0, $\sigma_s$=2.8-5.3~dB, PDP, CFR  \\
 \cline{2-7}
 & \cite{a5} &  15-100~m & WB, 2.585~GHz & Cellular & Suburban & $n$=1.3-3.7, $\sigma_s$=2.7/3.0~dB, PDP, KF, RMS-DS\\
\cline{2-7}
 & \cite{a6} &  15-120~m & NB, 800~MHz & Cellular & Suburban & $n$=2.0-3.7, $\sigma_s$=3.4-7.7~dB \\

\hline
\multirow{2}{*}{A2A}&  \cite{a7}  & 0-50~m & NB, 2.4~GHz & UAV & Suburban &  $n$=2.3-2.7, $\sigma_s$=1.9-5.5~dB \\
 \cline{2-7}
&\cite{a8}& 6-15~m & WB, 60~GHz& UAV & Urban & $n$=2.2-2.3, $\sigma_s$=0.8-8.0~dB  \\
\hline
\end{tabular}
 \begin{tablenotes}
        \footnotesize
        \item[]{Notes:} WB: Wideband, NB: Narrowband, SD: Stationarity distance, SIC: Spatial and inter-frequency correlation, CFR: Channel frequency response.
      \end{tablenotes}
\end{table*}

\subsection{Large-Scale Statistics}
Large-scale channel statistics such as path loss and shadow fading, are undeniably crucial for system design and performance evaluation. For the sake of a detailed description, we summarize the representative A2A and A2G channel measurements in Table~I, where important settings and corresponding channel statistics are included. Overall, we found that most measurements (A2A or A2G), focus on sub-6~GHz, where L-band, C-band, LTE band, and IEEE 802.11n band constitute the majority. Altitudes of aircraft range from several meters to hundreds of meters for A2G channels with GS terminal. For CC-UAV communications, UAV heights are lower than 120 m, owing to the limited power of TBS and specific regulations. For A2A channels, two UAVs generally fly in dozens of meters in existing measurements. As aforementioned, UAVs can be operated in various environments that were mostly considered in measurement campaigns.

Since path loss models are based on path loss exponent (PLE) and shadowing factor (SF), we focus on the two key parameters, where PLE and SF are denoted as $n$ and $\sigma_s$, respectively. It is found that as the altitude of UAV arises, the PLE becomes smaller for both A2A and A2G channels. As an example, in \cite{a4}, $n$ ranges from 2.5 to 3.0 for altitudes between 4 m and 16 m. Moreover, measurements in \cite{a1,a2} show that the values of $n$ are smaller than 2.2, which indicates that the A2G channel is close to free-space propagation for high altitudes. For CC-UAV communications, the PLEs vary in a large range as the height. For instance, the measurement in \cite{a6} at the LTE band, shows that $n$ can be up to 3.7 for near ground, which is similar to the result of traditional cellular channels. Notably, we found that the PLEs of A2A channels exhibit a certain relevance to the frequency. Measurements at 2.4 GHz in \cite{a7} show that $n$ gradually varies from 2.7 to 2.3 for the heights changing from 0 to 50 m. Whereas measurements at 60 GHz in low altitudes between 6-15 m show that $n$ is around 2.2-2.3 \cite{a8}. This phenomenon can be well explained by the density of MPCs. It is known that the MPCs can be dense in low frequencies due to the rich scattering, reflections, and diffractions, which contribute to a large PLE. However, these MPCs may be submerged under noise in mmWave bands, creating sparse MPCs and thus leading to a smaller PLE than that in low frequencies.


The shadow fading is generally modeled as a Gaussian random process with zero mean and standard deviation of $\sigma_s$. Table I shows that the typical values of SF for A2G channels range from 2.6-7.7 dB that is smaller than the typical values of 4-8.2~dB for cellular channels in outdoor environments suggested by the 3GPP. Besides, measurement in \cite{a7} shows the typical values of SF for A2A channels range from 1.9-5.5~dB, which are slightly lower than that in A2G channels since few scatters distribute in the vicinity of both UAVs. However, since measurements in \cite{a8} were carried out around several buildings, the SF can reach up to 8~dB. According to \cite{a4}, it is also found that the shadowing is highly related to the selected environment. The result shows that $\sigma_s$ is between 3.1 and 4.0~dB in an open-filed area, however, it becomes significant, ranging from 4.3 to 5.3~dB in a suburban environment.


\subsection{Small-Scale Statistics}
Small-scale channel properties are crucial to the related system design. For instance, the root-mean-square delay spread (RMS-DS) is used as a metric in the design of transmit symbols. Besides, the RMS-DS and Rician \emph{K}-factor (KF) are generally used to characterize the multipath propagation. Thus, we focus on the two representative parameters with analysis from the perspective of multipath effects.

In an over-sea environment \cite{a1}, average Rician KFs were reported to be 12.5 dB and 31.3 dB for L-band and C-band, respectively. Similar results were found for a suburban measurement in \cite{a2}, where the mean values of Rician KF were 12 dB and 27.4 dB for L-band and C-band, respectively. The similarity can be well explicated by the high altitude of the aircraft. Note that UAV flew above 504~m in both environments, which results in that the power of some of the multipath components (MPCs) caused by ground scatterers such as buildings become negligible, and thus these MPCs make a very limited contribution to the none-line-of-sight (NLOS) power. For low altitudes, the average Rician KF was 7.6~dB for 30-100 m reported in \cite{a5}, which illustrates the strong impact of terrestrial scatterers on the NLOS power.

The investigations of RMS-DS are as follows. In \cite{a1}, the median values of RMS-DS were 9.6-9.8~ns for numerous over-water measurements, whereas they concentrated around 9.6-11~ns for suburban and near-urban environments. A manifest difference lies in the largest RMS-DS, which generally occurs when buildings provide strong MPC reflections with long delays. For instance, the result indicates that the largest RMS-DS was 364.7~ns for an over-water environment, however, it reaches 4.24~$\mu$s in a suburban environment \cite{a2}.


\subsection{Channel Models}
Channel models provide formulated expressions that can be straightly applied in the system design and evaluation. Accordingly, we herein focus on discussing essential channel models including the path loss model, small-scale fading model, and MPC model for UAV communications. Path loss models are mainly composed of $\log$-distance \cite{a2}, modified free-space \cite{a4}, close-in (CI) \cite{a5}, floating intercept (FI) \cite{a6}, and excess loss (EL) \cite{a10} models. In these models, the $\log$-distance model is the most popular one for both A2A and A2G channels. Note that the EL model can be merely exploited in A2G channels since it separates the total loss into terrestrial and aerial parts. Moreover, the 3GPP model is tailored for CC-UAV communications, aiming to enable the LTE/5G network to provide communication services for aerial vehicles \cite{a11}.

The small-scale fading model mainly indicates the probability distribution of the envelope of small-scale fading and can be used to describe both narrowband fading channels and individual MPCs in wideband channels. Popular probability distributions consist of Rician, Rayleigh, and Nakagami-\emph{m} distributions. Abundant measurements have revealed that A2A and A2G channels are more likely to experience Rician fading because of the presentence of a strong LOS path in most cases \cite{a1,a2,a5}. Whereas Nakagami-\emph{m} can be a more general representation with the increasing multipaths, such as in \cite{a4}. Rayleigh fading is rare in existing works, nonetheless, research in \cite{a13} shows that A2G channels experience Rayleigh fading under NLOS condition in low altitudes.

Wideband measurements and simulations make it easier to detect MPCs and characterize their properties, which facilitates the reconstruction of the channel impulse response (CIR). An intuitive representation of the UAV channel resides in the two-ray propagation model. However, more MPCs were observed in measurements \cite{a1}, where a three-ray model was found to model A2G channels well for an over-water environment, even though the third ray is intermittent and needs to be expressed in a probabilistic way. Besides, over-three-ray models were found to model both A2A and A2G channels in built-up environments, where rays impinging on the walls of buildings result in more MPCs. It was found that MPCs for A2G channels can reach nine in a suburban environment \cite{a2}. Moreover, the latest A2A channel measurements and ray-tracing simulations in the reconstructed suburban environment have shown that the number of MPCs is up to seven, composed of the LOS, the ground reflection, four single-bounce wall reflections, and one double-bounce wall reflection \cite{a14}.

\section{Case Studies}
In this section, we conduct several case studies to further illustrate the UAV channel characterization and modeling. We will first show the impacts of frequency and environment on A2G channel characterizations based on numerous channel measurement campaigns. Then, we will employ an artificial neural network (ANN) to predict the path loss of A2A links.

\subsection{A2G Channel Measurements at 4 GHz and 24 GHz}
Since C-band and quasi-mmWave play a vital role in the high-data-rate payload communications of UAVs, we conducted A2G channel measurements at 4~GHz and 24~GHz in a campus environment. More specifically, a DJI drone equipped with a transmitter (Tx) flew vertically from 0 to 24~m, and a ground receiver (Rx) was placed on the rooftop of a building with a height of 25~m. Due to the blockage of a building, the channel experiences NLOS propagation for 0-11 m and LOS propagation for 11-24 m. Besides, the horizontal distance between Tx and Rx is around 350~m.

Path loss results with narrowband measurements are depicted in Fig.~2. It shows that path losses for both frequencies present similar trends concerning UAV heights. In particular, the PLEs based on the 3GPP path loss model \cite{a11} at 4/24~GHz are 2.42/2.49 and 3.79/3.76 for the LOS and NOLS case, respectively. However, it can be intuitively observed that the fluctuations of path loss at 24~GHz are more apparent in both LOS and NLOS cases than those at 4~GHz. Such fluctuations are mainly caused by small-scale fading, and thus can be quantitatively characterized by the fading depth (FD) that is defined as the difference in power values between the 50\% and 1\% level of small-scale fading and can be used to determine link fading margin. We found that FDs at 4~GHz and 24~GHz are 6.8~dB and 4.8~dB for the NLOS case, respectively, whereas for the LOS condition, FDs are 4.6~dB and 1.2~dB for 4~GHz and 24~GHz, respectively. The results demonstrate that A2G channels at high frequency are more vulnerable to suffering from deep fading.


\begin{figure}[!t]
  \centering
  \includegraphics[width=3.2in]{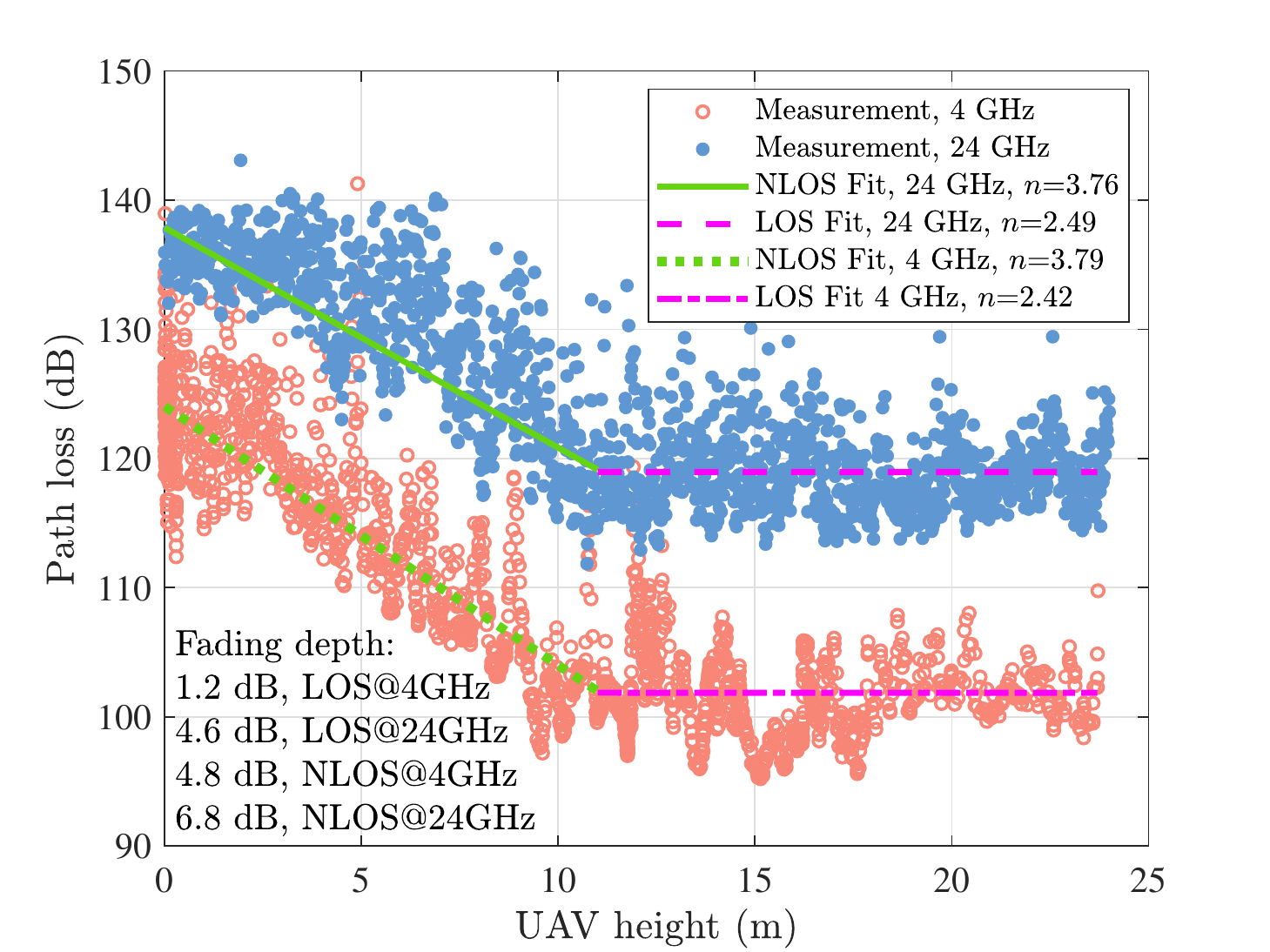}
  \caption{Path loss results with channel measurements at 4 GHz and 24 GHz.}
  \label{fresnel}
 \end{figure}
\subsection{A2G Channel Measurements in Lake and Suburban Areas}
Wideband measurements were conducted with two identical commercial DWM1001 modules. Each module includes an ultra-wideband (UWB) antenna, radio frequency circuitry, and a motion sensor. One module used as a Tx was mounted on the bottom of the six-rotor DJI drone, and the other was used as an Rx, placed at 0.5 m above the ground. UWB antennas for both ends are vertically polarised and approximately omnidirectional with 0 dBi gain. Moreover, the center frequency was at 6.5 GHz with a bandwidth of 500 MHz, providing a high delay resolution (2~ns) that is supportive of possibly capturing the most of MPCs. Due to the limited transmit power ($-$17 dBm), the test UAV flew in altitudes lower than 30~m and the horizontal distance between Tx and Rx is less than 95~m. Measurement campaigns were performed in two distinct scenarios, including over-lake and suburban environments.

\begin{figure}[!t]
  \centering
   \subfigure[Over-lake environment]{\includegraphics[width=3.2in]{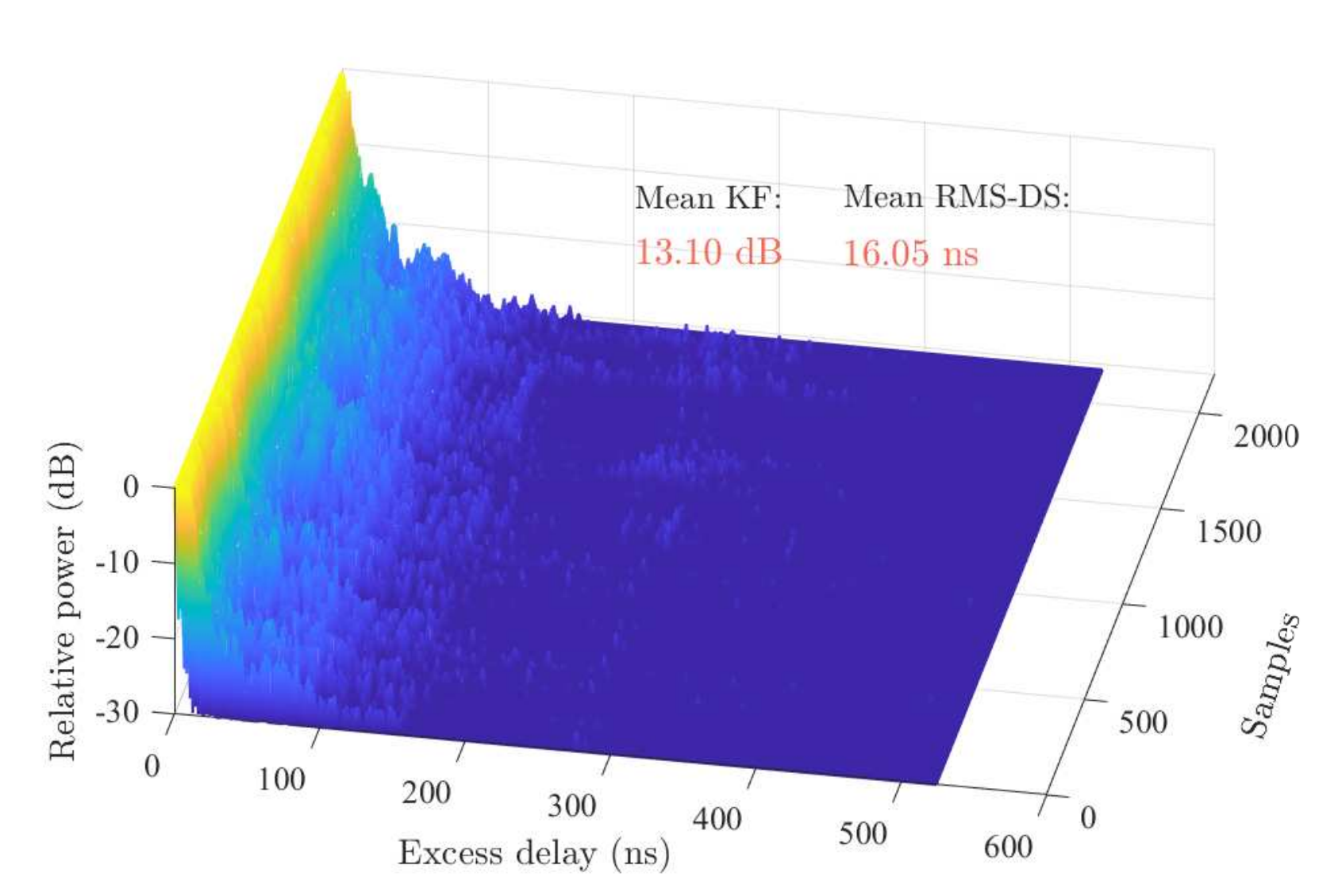}}
   \subfigure[Suburban environment]{\includegraphics[width=3.2in]{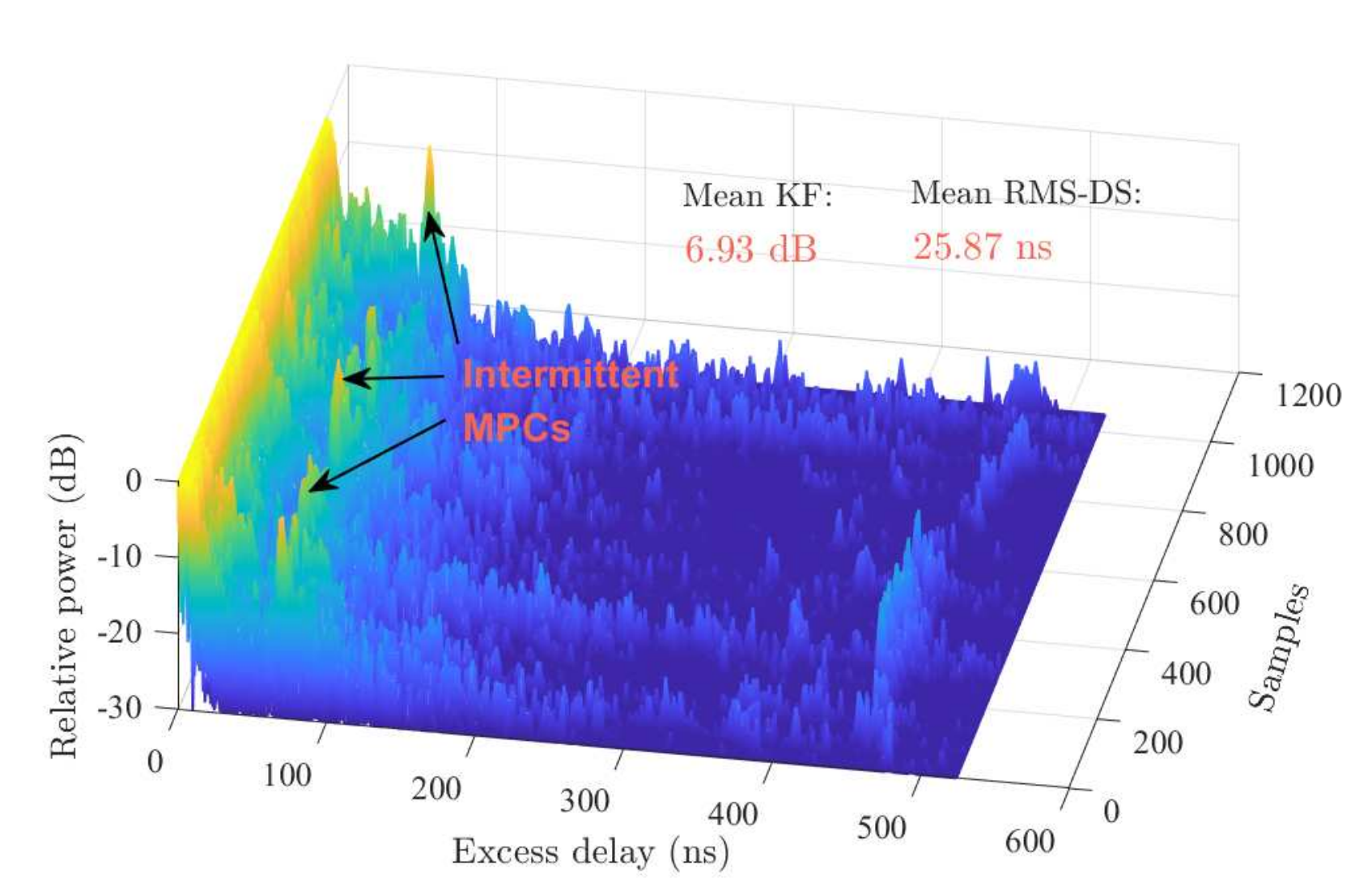}}
  \caption{A comparison of PDPs in different propagation environments.}
  \label{fresnel}
 \end{figure}

Numerous PDPs were obtained by wideband channel measurements, which enable to compare the discrepancy in terms of MPCs and essential channel parameters in different environments. For the post-processing of raw PDPs, the dynamic range remains as 30~dB, which implies that the MPCs with their power lower than 30 dB of the LOS power were neglected. Fig.~3 depicts processed PDPs for both environments. We can find that there are more intermittent MPCs in the suburban environment than that in the over-lake environment. The MPCs from buildings can favorably account for the divergence. Besides, a quantitative analysis was performed concerning small-scale channel parameters. We first extracted potential MPCs from PDPs using a high-resolution estimation algorithm, i.e., space-alternating generalized expectation-maximization (SAGE) algorithm. With extracted MPCs, we then calculated Rician KF and RMS-DS. As shown in Fig.~3, mean KFs are 13.10~dB and 6.93~dB for over-lake and suburban environments, respectively. For the RMS-DS, the mean values are 16.05~ns and 25.87~ns for over-lake and suburban environments, respectively. The large Rician KF, as well as small RMS-DS, illustrate the sparsity of MPC in an over-lake environment, which follows the result in \cite{a1} where a three-ray model was used to describe A2G channels. By contrast, MPCs present a denser distribution in the suburban environment. By means of SAGE, we found that the typical number of MPCs ranges from 7 to 12 in the suburban environment. The findings are beneficial for providing a more practical channel model.

\begin{figure}[!t]
  \centering
  \includegraphics[width=3.2in]{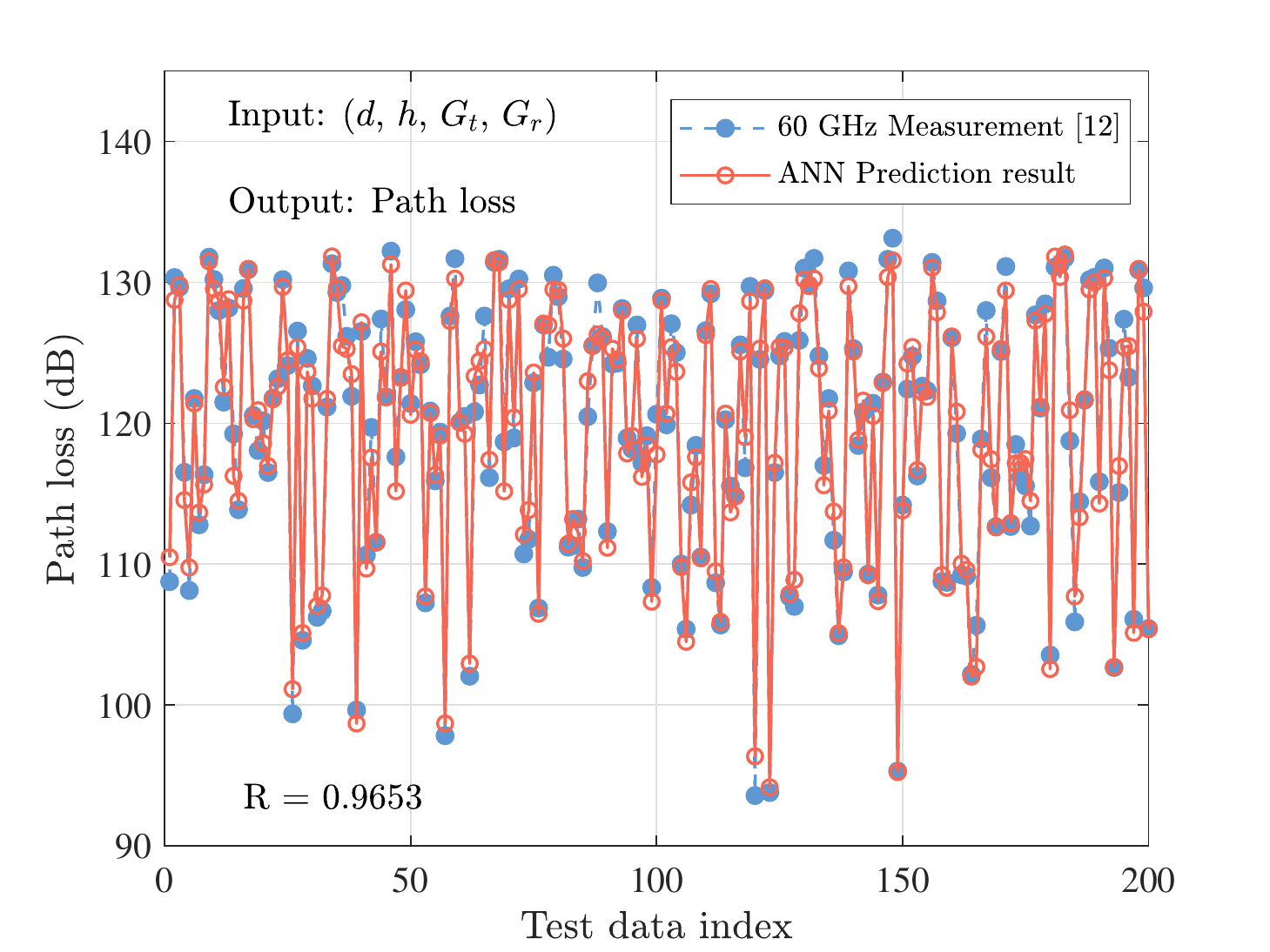}
  \caption{Results of path loss prediction by using artificial neural network.}
  \label{fresnel}
 \end{figure}

\subsection{Path Loss Prediction for A2A Channels via ANN}
Machine learning (ML), as a powerful prediction framework based on a comprehensive dataset and a flexible network architecture, has been widely employed in wireless communications. Meanwhile, ANN using a feed-forward structure and an assortment of ML algorithms can obtain more accurate predictions than standard empirical methods while being more computationally efficient than the deterministic method. Thus, the study aims to show that the ANN works well in aerial propagation channel predictions. Notably, raw training data from \cite{a8} was produced by extensive measurements for a UAV-to-UAV communication link. More specifically, measurements employed two DJI M600 UAVs at the same heights (6, 12, 15~m) and two Facebook Terragraph mmWave radios configured at 60 GHz as channel sounders with beam scanning capabilities. Moreover, the half-power beamwidth is $2.8^\circ$, the radio maximum effective radiated power is 45~dBm, and the horizontal distance between UAVs ranges from 6 to 40~m.

 For generality, few but necessary variables such as the horizontal distance ($d$), the height of UAVs ($h$), and the beam gains of Tx and Rx ($G_t$ and $G_r$) were chosen as the inputs of the network. The training trials show that more hidden layers do not increase the accuracy, thus, we use a single hidden layer to train. For the hidden layer, we also found three neurons are enough to achieve accurate predictions. The network is trained with the Levenberg-Marquardt backpropagation algorithm for updating the weights and biases of neurons. The activation function utilizes the commonly used hyperbolic tangent sigmoid function. Fig.~4 illustrates the accurate predictions of ANN for test data sets. In particular, the regression $R$ values measure the correlation between predictions and measurement results, with $R=0.9653$ showing the high accuracy.

\section{Future Directions}
In the former sections, we have summarized emerging challenges concerning UAV channel modeling, reviewed recent progress, and illustrated A2A and A2G channel characterizations and predictions based on case studies. In this section, we will point out prospective research directions, aiming to provide practical proposals for future works.

\subsection{UAV Channel Measurements that Fill the Current Gaps}
It is a perpetually continuous work for channel modeling study to consider new frequencies, environments, UAV heights, and antenna settings with the advent of the next-generation communication system. With 6G, new frequencies are considered, such as Terahertz (THz) and free-space optical bands. To our best knowledge, there is no existing UAV channel measurement work in these bands. Besides, more propagation environments can be involved in UAV channel measurements, compared with traditional built-up, over-water, forest, and mountain environments. For example, the outdoor-to-indoor (O2I) and indoor-to-outdoor (I2O) channel measurements are necessary for the design of aerial communications where the UAV serves as an ABS. Additionally, since current measurement campaigns mostly focus on low altitudes of UAV, measurements at higher altitudes are of importance to establish height-dependent channel models. Finally, there are limited measurements that were conducted for the multi-input multi-output (MIMO) antenna setting of UAV, which is not conducive to the characterization of the angular domain, such as angle of arrival (AoA) and angle of departure (AoD). Overall, there are still many gaps that need to be filled by measurements in emerging bands, more heights, multiple antennas, and distinct propagation environments.


\subsection{Machine Learning Empowered UAV Channel Modeling}
 Compared with conventional methods of channel modeling, ML-based methods have attracted much attention in recent years. We have shown the ability of ANN in predicting the path loss of the A2A channel in the case study. In addition to channel prediction, the ML can be accurate and efficient in the MPC clustering and scenario classification by novel learning algorithms. However, one of the challenges in ML-based channel modeling is that the learning process highly relies on the volume of the data set, which requires numerous measurements and simulations. Moreover, the designs of network architecture, feature extraction, and training parameters directly determine the accuracy and efficiency of the channel model. The last issue is concerning the generality of ML-based channel models. Since measurement and simulation data is mainly produced in the specific settings of frequency, environment, and communication links, whether the trained model works in the link with a new setting is uncertain and needs more validations in future researches.

\subsection{Channel Modeling for Cellular/Satellite-Connected UAVs}
The space-air-ground integrated network is regarded as a solution to realize global coverage in 6G, where UAV, as a key enabler in the air (middle) layer, plays an important role in connecting both satellite and ground cellular networks. Thus, satellite-to-air (S2A) and air-to-BS channels are worthy of much attention. We have highlighted the key challenge with the directional beam of TBS with down-tilt sector antennas in CC-UAV communications. Besides, the interferences from peripheral TBSs are significant due to the large LOS probability for high-altitude UAVs. For S2A links, although there may be no scatterers in the vicinity of both ends, atmospheric effects, ground-based originated scattering, satellite elevations, and local effects at the aircraft structure itself can lead to severe fading of the received signal amplitude, which yields a potential direction for in-depth study.


\subsection{Quantifying Impacts of UAV Internal Factors on Channels}
In existing A2A and A2G channel modeling, the terminals are generally assumed as non-physical nodes without premeditating their sizes. However, the experiences from vehicular channel modeling remind us that the physical size of the vehicle has a significant impact on channel properties. As we mentioned before, the large-scale shadowing and small-scale fading caused by the UAV internal factors are worth considering. For example, the jittering produced during the flight of UAV has been investigated for its impact on the delay spread of A2G channels \cite{f2}. It is confirmed that a more detailed channel study is becoming increasingly popular with comprehensive considerations of the physical structure of UAV, the stability of flight, the performance of the engine, the endurance of battery, and other exhaustive details. However, it is challenging to quantify the impacts of these factors, which not only needs massive measurements but also requires rigorous theoretical derivations.

\section{Conclusion}

In this article, we focus on A2A and A2G propagation channel modeling and characterization for UAV communications. We first summarized key challenges for UAV channel modeling in terms of the frequencies, environments, as well as the size, altitude, mobility, and flying state of UAV. Afterward, a comprehensive survey of the state of the art clearly illustrates the large-scale and small-scale statistics of A2A and A2G channels. We also provided a summary of channel models such as path loss, small-scale fading model, and multipath model, where the $\log$-distance path loss model and Rician distribution are the most commonly used to describe the large-scale and small-scale fading, respectively. We subsequently conducted measurement-based case studies to show the impacts of frequency and environment on A2G channel characterizations. We also carried out the case study to show the accuracy of path loss prediction for A2A channels through machine learning methods. Finally, we provided practical and prospective directions, concerning indispensable measurements, novel methodologies, emerging scenarios, and quantifiable influences of UAV sizes, structures, and flights on channel properties.

\section*{Acknowledgement}
The authors would like to acknowledge support from the Key-Area Research and Development Program of Guangdong Province, China (2019B010157001), NSFC under Grant (61771036, 61901029), the State Key Laboratory of Rail Traffic Control and Safety (Contract No. RCS2020ZZ005). The first author would like to thank the financial support from the China Scholarship Council (CSC) (Grant No. 202007090173). The corresponding author of this article is Ke Guan.

\newpage

\begin{IEEEbiographynophoto} {Zhuangzhuang Cui} [S'18] (cuizhuangzhuang@bjtu.edu.cn) received the B.S. degree in communications engineering from Beijing Information Science \& Technology University, Beijing, China, in 2016. He is currently pursuing a Ph.D. degree with the State Key Laboratory of Rail Traffic Control and Safety, Beijing Jiaotong University, Beijing, China.
\vspace{-220 pt}
\end{IEEEbiographynophoto}

\begin{IEEEbiographynophoto}{Ke Guan} [SM'20] (kguan@bjtu.edu.cn) received the B.S. degree in communications engineering and Ph.D. degree from Beijing Jiaotong University (BJTU), Beijing, China, in 2006 and 2014, respectively. Since 2020, he is a Full Professor with BJTU. He was a recipient of the 2014 URSI Young Scientist Award. His papers received IEEE Vehicular Technology Society 2019 Neal Shepherd Memorial Best Propagation Paper Award.
\vspace{-220 pt}
\end{IEEEbiographynophoto}

\begin{IEEEbiographynophoto}{C\'esar Briso-Rodr\'iguez}[M'00] (cesar.briso@upm.es) received the Ph.D. degree in mobile communications from the Universidad Polit\'ecnica de Madrid (UPM), in 1998. He is currently a Full Professor with the UPM. He has made 20 industrial and research projects with national and international companies and institutions.  He was the Recipient of the National Awards for the Best Ph.D. of the Spanish Association of Telecommunications Engineers.
\vspace{-220 pt}
\end{IEEEbiographynophoto}

\begin{IEEEbiographynophoto}{Bo Ai} [SM'10] (boai@bjtu.edu.cn) received his M.S. and Ph.D. degrees from Xidian University, China, in 2002 and 2004, respectively. He was a Visiting Professor with the Electrical Engineering Department, Stanford University, in 2015. He is a Full Professor and the deputy director at the State Key Laboratory of Rail Traffic Control and Safety at Beijing Jiaotong University. He is an IET Fellow and an IEEE VTS Distinguished Lecturer.
 \vspace{-220 pt}
\end{IEEEbiographynophoto}

\begin{IEEEbiographynophoto}{Zhangdui Zhong} [SM'16] (zhdzhong@bjtu.edu.cn) received the B.S. and M.S. degrees from Beijing Jiaotong University, Beijing, China, in 1983 and 1988, respectively. He is currently a Full Professor and Chief Scientist of the State Key Laboratory of Rail Traffic Control and Safety. He was a recipient of the MaoYiSheng Scientific Award of China, the ZhanTianYou Railway Honorary Award of China, and the Top ten Science/Technology Achievements Award of Chinese Universities.
\vspace{-220 pt}
\end{IEEEbiographynophoto}

\begin{IEEEbiographynophoto}{Claude Oestges} (F'16) (claude.oestges@uclouvain.be) received the M.Sc. and Ph.D. degrees in electrical engineering from the Universit\'e catholique de Louvain (UCLouvain), Louvain-la-Neuve, Belgium, in 1996 and 2000, respectively. He is currently a Full Professor with the Institute of ICTEAM, UCLouvain. Dr. Oestges has been the Chair of COST Action CA15104 IRACON, since 2016. He was a recipient of the IET Marconi Premium Award in 1999 to 2000 and the IEEE Vehicular Technology Society Neal Shepherd Awards in 2004 and 2012.

\end{IEEEbiographynophoto}

\end{document}